\documentstyle[12pt,epsf]{article}
\newcommand{\be}{\begin{equation}}
\newcommand{\ee}{\end{equation}}
\newcommand{\ba}{\begin{eqnarray}}
\newcommand{\ea}{\end{eqnarray}}
\newcommand{\bb}{\begin{thebibliography}}
\newcommand{\eb}{\end{thebibliography}}
\newcommand{\ci}[1]{\cite{#1}}
\newcommand{\bi}[1]{\bibitem{#1}}
\newcommand{\lab}[1]{\label{#1}}

\newfont{\prg}{cmsy10}

\begin{document}

\begin{center}
{\bf
  Behavior of the Hadron Potential at Large Distances \\
  and Properties of the Hadron Spin-flip Amplitude     }
\vskip 5mm

  E. Predazzi$^{1}$,   O.V. Selyugin$^{2}$
\vskip 5mm
{\small
 {\it
 $^{1}$Dartimento di Fisica Teorica - Unversit\`{a} di Torino
   and Sezione INFN di Torino, Italy \\
  $^{2}$BLTPh, JINR, Dubna, Russia         }}
\end{center}
\vskip 5mm

\begin{abstract}
  The impact of the form of the hadron potential  at large distances
  on the behaviour of the hadron spin-flip amplitude at small angles
  is  examined.
  The $t$-dependence of the spin-flip amplitude of  high energy
  hadron elastic scattering is analyzed under different assumptions
  on the hadron interaction.
  It is shown that the long tail
  of the non-Gaussian form of the hadron potential
   of the hadron interaction in the impact parameter
  representation leads to a large value of the slope of the spin-flip
   amplitude (without the kinematical factor $\sqrt{|t|}$) as compared with the
  slope of the spin-non-flip amplitude.
  This effect can explain
   the form of the differential cross section
  and the analyzing power at small transfer momenta.
   The methods for the definition of the spin-dependent
   part of the hadron scattering amplitude are presented.
  A possibility to investigate the structure of the hadron spin-flip amplitude
  from the accurate measure of the differential cross section and the
  spin correlation parameters is shown.
 \end{abstract}

\vskip 10mm

\section{Introduction}

    Spin phenomena provide a powerful tool for analyzing
    the  properties  of  hadronic interaction.
  The spin structure of the pomeron is an important question
 concerning the  diffractive scattering of polarized particles.
There were many observations of spin effects at high energies and
at fixed momentum transfers.
 Several
attempts to extract the spin-flip amplitude from the experimental
data  show that the ratio of spin-flip to spin-nonflip
amplitudes can be non-negligible and may be only slightly dependent of energy
\cite{akch,sel-pl}.
   Thus, the diffractive polarized experiments at HERA and RHIC
 allow one to study spin properties of quark-Pomeron and
 proton-Pomeron vertices
 and  to search  for a possible  odderon contribution.

      This provides an important test of the
 spin properties of QCD at large distances.
In all of these cases,  pomeron exchange
is expected to contribute the observed spin effects at some level
 \cite{martini}.

In the framework of perturbative QCD,
 it was shown that
the analyzing power of hadron-hadron
scattering  can be large and proportional to
 the hadron mass \cite{ter}.  Hence, one could
expect a large analyzing power for moderate $p_{t}^{2}$
where the spin-flip amplitudes are presumably relevant
 for  diffractive processes.
The case,  when large distance
contributions are considered, leads to a more complicated
spin structure of the pomeron coupling.
For example, the spin-flip amplitude has been estimated in the QCD Born
approximation by using the nonrelativistic quark model for the
nucleon wave function \cite{kopel} in the case where the nucleon
contains a dynamically enhanced component with a compact diquark.
The spin-flip part of the scattering
amplitude can be determined by the hadron wave function for the
pomeron-hadron couplings or by the gluon-loop corrections for the
quark-pomeron coupling \cite{gol-pl}.  As a result, spin asymmetries
appear which have a weak energy dependence as $s \to \infty$.
Additional spin-flip contributions to the quark-pomeron vertex may
also have their origins in instantons (see {\it e.g.} \cite{fo,do}).

   The procedure of how to separate the various  parts
 of the scattering amplitude is the model-dependent.
   These questions include  also
   the determination of the odderon contribution to  different
   exclusive reactions and to  pseudoscalar meson production,
   the study of the structure of high energy elastic
   hadron-hadron scattering amplitude at small angles
    and the problems
   related to the extraction of $\sigma_{tot}$ from the experimental data,
   the study of the behavior of the parameter $\rho$, -
   the ratio the real to imaginary part of scattering amplitude
   \cite{str,osc}.

 Including in the analysis the experimental data of spin correlation
parameters does not simplify the task.
 In the general case,
the  form of the analyzing power, $A_N$, for example, and
 the position of the maximum of  $A_N$
 depends on the parameters of the elastic scattering
 amplitude $\sigma_{tot}$,  $\rho(s,t)$, the Coulomb-nucleon interference
 phase  $\varphi_{cn}(s,t)$
 and the elastic slope $B(s,t)$ .
 For the definition of new effects at small angles
  and especially in the region of the diffraction minimum
  one must  know the effects of the Coulomb-hadron interference
 with sufficiently high accuracy.
  The Coulomb-hadron phase was calculated
 in the entire diffraction domain taking into account  the form factors
 of the  nucleons \cite{prd-sum}.
  Some polarization effects connected with the Coulomb hadron
   interference, including some possible  odderon contribution,
   were also calculated \cite{z00}.

   The dependence of the hadron spin-flip amplitude on the momentum transfer
  at small angles is tightly connected with the basic structure of the
  hadrons at large distances. We show that
  the slope of the so-called ``reduced'' hadron spin-flip amplitude
  (the hadron spin-flip amplitude without the kinematic factor $\sqrt{|t|}$)
  can be larger
  than the slope of the hadron spin-non-flip amplitude
  as was observed long ago \cite{predaz,wak}.
  Its behavior can be
  defined by  small effects in the differential hadron cross section
  and  the real part of hadron non-flip amplitude.

\section{Definition of the physical quantities.}

 The total helicity amplitudes can be written
 as $ \Phi_i(s,t) = \phi^h_{i}(s,t)
        + \phi_{i}^{em}(t) \exp[i \alpha_{em} \varphi_{cn}(s,t)]$,
 where $\phi^h_{i}(s,t)$ is the pure strong interaction of hadrons,
  $\phi_{i}^{em}(t)$  is the electromagnetic interaction of hadrons,
  $\alpha_{em}=1/137$ is the electromagnetic constant,
   and
  $\varphi_{cn}(s,t)$ is the electromagnetic-hadron interference phase factor.
  So, to determine the hadron spin-flip amplitude
 at small angles
  one should take
  into account all electromagnetic and  interference
  electromagnetic-hadrons effects.
   A recent reanalysis of the data is found  in \cite{PPP}.

  The differential cross
  section and the spin parameters $A_N$ and $A_{NN}$ are defined as
 \ba
  \frac{d\sigma}{dt}&=& \frac{2 \pi}{s^2}(|\phi_1|^2+|\phi_2|^2+|\phi_3|^2
   +|\phi_4|^2+4|\phi_5|^2),
								\lab{dsdt}
 \ea
 \ba
  A_N\frac{d\sigma}{dt}&=& -\frac{4\pi}{s^2}
                 Im[(\phi_1+\phi_2+\phi_3-\phi_4) \phi_5^{*})],  \lab{an}
\ea
and
 \ba
  A_{NN}\frac{d\sigma}{dt}&=& \frac{4\pi}{s^2}
                [ Re(\phi_1 \phi_2^{*} - \phi_3 \phi_4^{*})
                + 2 |\phi_5|^{2}],  \lab{ann}
\ea
 in terms of the usual helicity amplitudes.


   In this paper we define the hadronic spin-nonflip amplitude as
  $F^{h}_{nf}(s,t)
    = (\phi^h_{1}(s,t) + \phi^h_{3}(s,t))/(2s)$
  and
 $F^{c}_{nf}(s,t)
    = (\phi^{em}_{1}(s,t) + \phi^{em}_{3}(s,t))/(2s)$
 Taken into account the Coulomb-nuclear phase $\varphi_{cn}$ we
 define $Im F_{nf}^{c} = \alpha_{em} \varphi_{cn} F_{nf}^{c}$.

The   ``reduced''    spin-flip amplitudes are denoted as
  $\tilde{F^{h}_{sf}}(s,t) =  \phi^{h}_{5}(s,t)/(s \sqrt{|t|})$  and
  $\tilde{F^{c}_{sf}}(s,t) =  \phi^{em}_{5}(s,t)/(s \sqrt{|t|})$.

 \section{The slope of the hadron  amplitudes}

  As it is not possible to caculate  exactly  the hadronic  amplitudes
  from  first principles,
  we have to resort to some assumptions for what concerns
  their form ($s$ and $t$  dependence).
   Let us define the slope of the scattering amplitude as
   the derivative of the logarithm of the amplitudes with respect to $t$.
     For an exponential form of the amplitudes this  coincides
    with the usual slope of the differential cross sections divided by $2$.

  If we define the forms of the separate  hadron scattering
  amplitude as:
\ba
Im \ F_{nf}(s,t) \sim exp(B_{1}^{+} \ t), \ \ \
  Re \ F_{nf}(s,t) \sim exp(B_{2}^{+} \ t),
\ea
\ba
 Im  \tilde{F_{sf}}(s,t)=
 \frac{1}{\sqrt{|t|}} Im \phi^{h}_{5}(s,t) \sim  \ exp(B_{1}^{-} \ t), \ \ \
  Re \tilde{F_{sf}}(s,t) \  = \
\sim  \ exp(B_{2}^{-} \ t),
\ea
 then, at small $t$ ($\sim 0 \div 0.1 \ GeV^2$), practically all
 semiphenomenological analyses assume:
 $$ B_{1}^{+} \ \approx \ B_{2}^{+} \ \approx \
    B_{1}^{-} \ \approx \ B_{2}^{-} . $$

    Actually, if we take the eikonal representation for the scattering
   amplitude
\ba
\phi^{h}(s,t) = \frac{1}{2 i \pi} \int d^{2} \rho
  e^{i \ \vec{\rho} \ \vec{\Delta}}  \
 [e^{\chi_{0} \ + \ i[\vec{n} \times \vec{\sigma}]_{z} \chi_{1}} \  - \ 1 ],
\ea

and use
$$ J_{0}(x) = \frac{1}{2\pi} \int_{0}^{2\pi} d\vartheta e^{i x cos \vartheta};
\ \ \
  J_{1}(x) = -\frac{1}{2\pi} \int_{0}^{2\pi} d\vartheta e^{i x cos \vartheta}
	 \sin{\vartheta}, $$
 we obtain
\ba
\phi^{h}_{1}(s,t) = - i p \int_{0}^{\infty} \ \rho \ d\rho
 \ J_{0}(\rho \Delta)
 (e^{\chi_{0}(s,\rho)} \  - \ 1 ],
\ea
\ba
\phi^{h}_{5}(s,t) = - i p \int_{0}^{\infty} \ \rho \ d\rho
 \ J_{1}(\rho \Delta) \ \chi_{1}(s,\rho) \ e^{\chi_{0}(s,\rho)} .
\ea
where
\ba
\chi_{0}(s,\rho) = \frac{1}{2ip} \int_{-\infty}^{\infty} dz
 \   V_{0}(\vec{\rho}, z);
\ea
\ba
\chi_{1}(s,\rho) = \frac{\rho}{2} \int_{-\infty}^{\infty} dz
  \ V_{1}(\vec{\rho}, z)
\ea

  If the potentials $V_{0}$ and $V_{1}$ are
   assumed to have a Gaussian form
$$ V_{0,1}(\rho, z) \sim \ \int_{-\infty}^{\infty} e^{B \ r^2} \ d z
 \ = \ \frac{\sqrt{\pi}}{\sqrt{B}} e^{-B \ \rho^2} $$
  in the first Born approximation
  $\phi^{h}_{1}$ and $\hat{\phi_h}^{5}$
   will have  the same form
\ba
\phi^{h}_{1}(s,t) \sim  \int_{0}^{\infty} \ \rho \ d\rho
 \ J_{0}(\rho \Delta) e^{-B \ \rho^2} \ = \ e^{-B \Delta^{2}}, \lab{f1a}
\ea
\ba
\phi^{h}_{5}(s,t) \sim \int_{0}^{\infty} \ \rho^2 \ d\rho
 \ J_{1}(\rho \Delta) \  e^{\chi_{0}(s,\rho)}
   \ e^{-B \ \rho^2 } \ = \ q \ \ B \ e^{-B \Delta^{2}}  . \lab{f5a}
\ea

  In this special case, therefore,
the slopes of
 the  spin-flip and  ``residual''spin-non-flip amplitudes are
  indeed the same.

  At best, however, a Gaussian form of the potential
 is adequate to represent   the central part of the
   hadronic  interaction. This form cuts off the Bessel function
  and the contributions at large distances.
 As a consequence
  we can confine ourselves to the two first terms of
 the small-$x$ expansion of the Bessel functions
\ba
 J_{0}(x) \ \simeq \ 1 \ - \ (x/2)^2; \ \ \ and \ \ \
  2 \ J_{1}/x \ = \ (1 \ - 0.5 \ (x/2)^2) , \lab{sbes}
\ea
  and we recover the previous result, i.e.,
 the integral representation   for spin-flip and spin-non flip amplitudes
 will be the same as in (\ref{f1a},\ref{f5a}).
  If, however, the potential (or the corresponding eikonal)
 has a long tail (exponential or power)
  in the impact parameter,  
  the approximation (\ref{sbes}) for the Bessel functions
  does not lead to a correct results and one has to perform
 the full integration.

  The first observation that the slopes don't coincide
 was made in \cite{predaz}.
  It was found from the analysis of the
 $\pi^{\pm} p \rightarrow \ \pi^{\pm}p $ and
  $pp \rightarrow \ pp $
reactions
  at $p_L \ = \ 20 \div 30 \ GeV/c $
  that the slope of the ``residual'' spin-flip amplitude is about
   twice as large as
  the slope of the spin-non flip amplitude. This conclusion can also
  be obtained
  from the phenomenological analysis carried out in \cite{wak} for
  spin correlation parameters of the elastic proton-proton scattering
  at $p_L \ = \ 6 \ GeV/c$.

   The model-dependent analysis based on all the existing experimental
   data of the spin-correlation parameters above $p_L \ \geq
 \  6 \ GeV$
   allows  us to determine the structure of the hadron spin-flip
   amplitude at high energies and to predict its behavior at
   superhigh energies \cite{yaf-wak}. This analysis shows
   that the ratios
   $Re \ \phi^{h}_{5}(s,t) / (\sqrt{|t|} \ Re \ \phi^{h}_{1}(s,t))$ and
   $Im \ \phi^{h}_{5}(s,t)/(\sqrt{|t|} \ Im \ \phi^{h}_{1}(s,t))$
   depend on $\ s$ and $t$. At small momentum transfers,
   it was found that the slope of the ``residual'' spin-flip
    amplitudes is approximately
   twice the slope of the spin-non flip amplitude.

    Let us see what we obtain in the case of an exponential tail
  for the  potentials.
 If we take
$$ \chi_{i}(b,s) \ \sim \ H \ e^{- a \ \rho}, $$
 and use the standard integral representation
\ba
\int_{0}^{\infty} \ x^{\alpha-1} - exp(-p \ x) J_{\nu}(cx)\ dx \
   = \ I_{\nu}^{\alpha}, \nonumber
\ea
with
\ba
 I_{\nu}^{\nu+2} \ = \ 2 p \ (2c)^{\nu} \ \Gamma (\nu + 3/2)
 1/[\sqrt{\pi} (p^{2}+c^{2})^{3/2}], \nonumber
\ea
we obtain
\ba
 F_{nf} (s,t) &=& \int \rho \ d\rho \ e^{- a \ \rho }
  \ J_0 (\rho q) \nonumber \\
  &=& \frac{a}{(a^2 + q^2)^{3/2}} \approx \
 \frac{1}{a\sqrt{a^2+q^2}} \ e^{-B q^2} \nonumber
\ea
with $B \ = \ 1/a^{2}$,
 where we have used the
$$ 1/(1+x) \ \sim  \ (1-x) \ \sim \ exp(-x). $$
 For the ``residual'' spin-flip amplitude,
  on the other hand,
we obtain
\ba
\sqrt{|t|} \tilde{F_{sf}}(s,t) &=&
  \int \rho^{2} \ d\rho \ e^{- a \ \rho} \ J_1 (\rho q)  \nonumber \\
 &=& \frac{3 \ a \ q}{(a^2 + q^2)^{5/2}} \ \approx
 \ \frac{3 \ a  q \ B^{2}}{ \sqrt{a^2+q^2}} \ \ e^{-2 \ B q^2}
\ea
  In this case, therefore,
  the slope of the ``residual'' spin-flip amplitude exceeds the slope
 of the spin-non-flip amplitudes by a factor of two.

If, further, we take the first Born term of an inverse power
   form factor
\ba
\chi (s,\rho) &=& \int q \ dq \ J_0 (\rho q)
 \frac{\Lambda^{2n}}{(\Lambda^2 + q^2)^{n}} \
  = \ \frac{\Lambda^{2n}}{48}(\Lambda \rho)^{n-1} K_{n-1}(\Lambda \rho),
\ea
 the spin-flip amplitude  is given by
\ba
\sqrt{|t|}\tilde{F_{sf}}(s,t) &=&\frac{\Lambda^{2}}{48}
  \int \rho^{2} \ d\rho \  J_1 (\rho q)
       K_{n-1}(\Lambda \rho) \nonumber \\
 &=& \frac{8 \ q}{\Lambda^{2}} \
 \ \frac{1}{ (1 + q^2/\Lambda^2)^{n+1} } \nonumber
\ea

  Hence, a  long tail hadron potentials
 implies a
 significant difference of the
  slopes of the ``residual'' spin-flip and of the spin-nonflip amplitudes.
   Note also that the procedure
  of  eikonalization will lead to a further increase
  of the difference of these two slopes.

\section{The determination of the structure of the hadron spin-flip amplitude}

   How one can find the value of the slope of the spin-flip
  amplitude?  First, note that if the ``reduced'' spin-flip amplitude
  is not small,  the impact of a large $B^{-}$ will reflect
   in the behavior of the differential cross section at small
   angles \cite{sel-sl}.
  Of course, this method gives  only  the absolute value of the coefficient
  of the  spin-flip amplitude. The imaginary
  and real parts of the spin-flip amplitude can be found only from
  the measurements of the spin correlation coefficient.

 Now let us examine the form of the analyzing power $A_N$ at
 small transfer momenta with different values of the slope of
 the hadron spin-flip amplitude. To this aim, we  take the spin nonflip
 amplitude in the standard exponential form with
  definite parameters:
 slope $B^{+}$, $\sigma_{tot}$ and $\rho^{+}$. For the ``residual''
 spin-flip amplitude, on the other hand,
  we consider  two possibilities:  equal slopes
 $B^{-}=B^{+}$ and  $B^{-}= 2 B^{+}$.
 For example we take at $\sqrt{s} \  = \ 50 \ GeV$
$$\sigma_{tot} \ = \ 42 \ mb; \ \ \ \rho \ = \ 0.075; \ \ \ B^{+} \ =
 \ 13/2 .$$
 The results of these  two different
 calculations are shown in Fig.1. It is clear that around
 the maximum of the Coulomb-hadron interference, the difference between the two
 variants is very small. But when $|t| \ > 0.01 \ GeV^2$,
  this difference grows. So, if we  try to find the
  contribution of the pomeron spin-flip, we should take into account
  this effect.

\vskip -10mm

\epsfysize=9.cm
\epsfxsize=12.cm
\centerline{\epsfbox{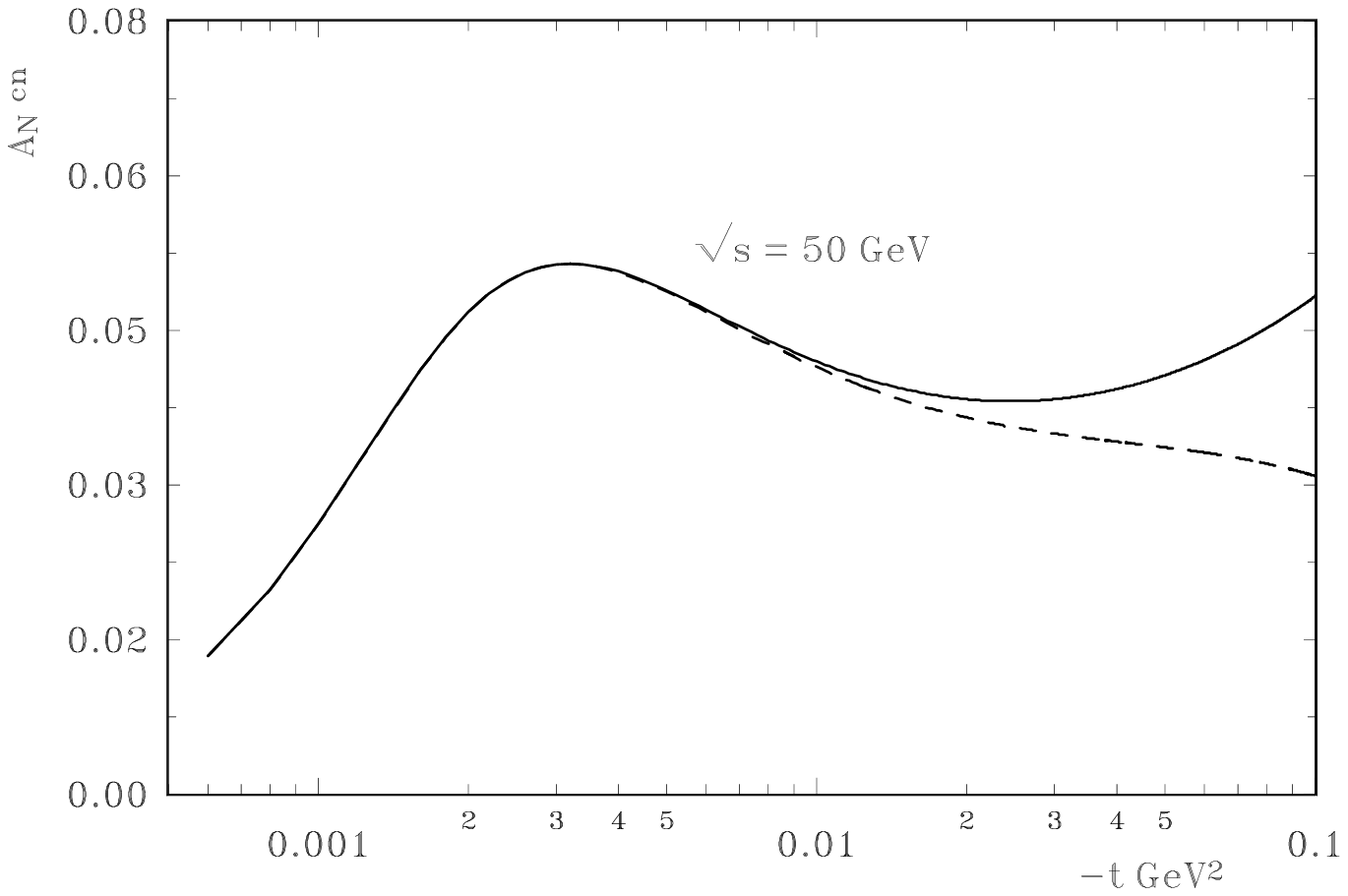}}

\vspace*{-5mm}

Fig. 1. \hspace{1mm}  The  calculation of $A_N $ at $\sqrt{50} \ GeV$
 (the solid line is
 with the slope $B_{1}^{-}$  of $\tilde{F_{sf}} $ equal to the slope
  $B_{1}^{+}$ of $F_{nf}$);
  the dashed line is
 with the $B_{1}^{-} \ = \ 2 \ B_{1}^{+} $ .

\vspace*{5mm}

As the value of $A_N$ depends on the determination of the beam
 polarization, let us calculate  the derivative of  $A_N$ with respect to $t$,
 for example, at $\sqrt{s} = 500 \ GeV$. In this case we take
 $$\sigma_{tot} \ = \ 62 \ mb; \ \ \ \rho \ = \ 0.15; \ \ \ B^{+} \ =
 \ 15.5/2 .$$

 The results of these  two
 calculations are shown in Fig.2.

\vskip -10mm

\epsfysize=9.cm
\epsfxsize=12.cm
\centerline{\epsfbox{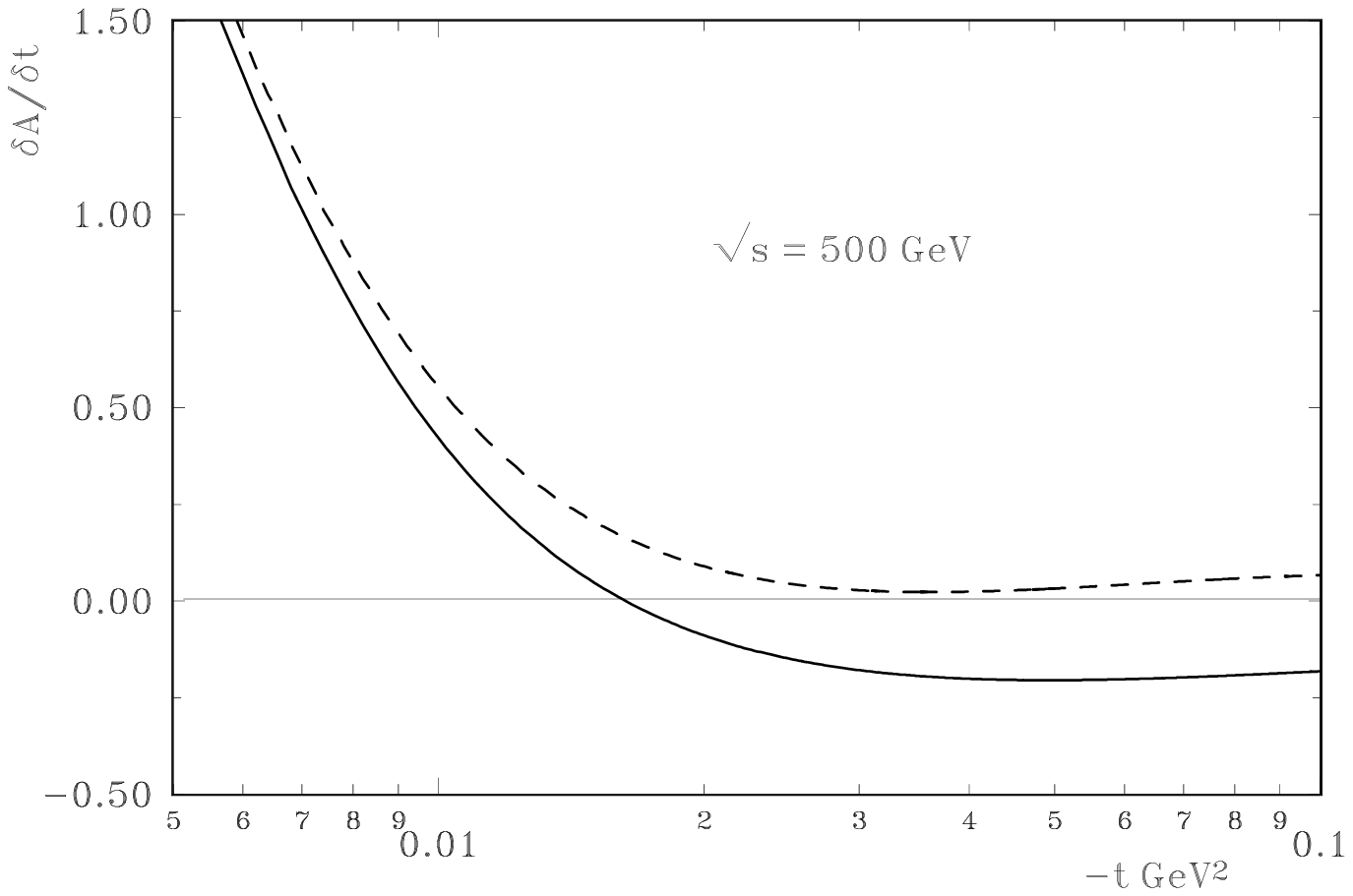}}

\vspace*{-5mm}

Fig. 2. \hspace{1mm}  The  calculation of $- \delta A_N/\delta t$
  at $\sqrt{500} \ GeV$:
   the solid line refers to  the case when
 the slope $B_{1}^{-}$  of $\tilde{F_{sf}} $ is equal to the slope
  $B_{1}^{+}$ of $F_{nf}$;
  the dashed line when
   $B_{1}^{-} \ = \ 2 \ B_{1}^{+} $ .

\vspace*{5mm}

   If we know the parameters of the hadron spin non-flip amplitude,
 the measurement of the analyzing power at small transfer momenta
 helps us to find the structure
 of the hadron spin-flip amplitude.
  Let us examine the behavior of the analyzing power (\ref{an}),
  which can be rewritten as
\ba
  A_N\frac{d\sigma}{dt}& =&
  2 (ImF_{nf}ReF_{sf}-ReF_{nf}ImF_{sf}) \\ \nonumber
 &&=  2  [( ImF_{nf}^h ReF_{sf}^c +ImF_{nf}^c ReF_{sf}^c
        - ReF_{nf}^h ImF_{sf}^c                          \\ \nonumber
   && -ReF_{nf}^c ImF_{sf}^c )
     + (ImF_{nf}^h ReF_{sf}^h -ReF_{nf}^c ImF_{sf}^h    \\ \nonumber
     & &
        +  ImF_{nf}^c ReF_{sf}^{h} -Re F_{nf}^{h} Im F_{sf}^{h})] \lab{anf}
\ea
  in two specific regions.

 We  begin from the point  $t_{im}$ where the absolute value of the
 real part of the Coulomb amplitude is
  equal  to the imaginary part of the hadron spin-non-flip amplitude
  $|Re F^{c}_{nf}| = |Im F_{nf}^{h}|$ .
 Let us denote by
  $P^{'}$  the part of the analyzing power (\ref{anf}) which can be calculated, if
   we know  $\sigma_{tot}, B, \rho$ for the non-flip amplitude, and
 by   $\Delta P$  the part of $A_N$ which depends
  on the hadron spin-flip amplitude.
\ba
  \Delta P = A_{N}^{exper.} -P^{'}.
\ea
Here $A_{N}^{exper.}$ is the measured value of the analyzing power.
Let us factor out the term  $|F_{nf}^c|$ from the numerator,
the term  $|F_{nf}^c|^{2}$ from the denominator and multiply
 $\Delta P$ by $|F_{nf}^c|$ :
\ba
  A_{N}^{'}= (\Delta P ) |F_{nf}^c| \simeq 2 \frac{-ImF_{sf}^h
      + ReF_{sf}^h (ImF_{nf}^h/|F_{nf}^c|}
     {1 + (ImF_{nf}^{h}/|F_{nf}^c|)^{2}} .
\ea
  At the point $t_{im}$ where $ ImF_{nf}^h = - ReF_{nf}^c$
  for  proton-proton scattering at high energy we  obtain
\ba
   A_{N}^{'} = - Re F_{sf}^h - ImF_{sf}^h .
\ea
 And, as $t \rightarrow 0$,due to the growth of $|F_{nf}^{c}|$
 we have the result
\ba
   A_{N}^{'} \rightarrow -2 ImF_{sf}^h.
\ea

  There is another specific point of  the differential cross sections
  and of $A_N$ on the axis of the momentum transfer, - $t_{re}$,
 where the  absolute value of the real part of the Coulomb amplitude  equals
   the absolute value of the real part of the hadron spin-non flip amplitude.
  This point $t_{re}$  can be found from
  the  measurement of
 the   differential cross sections   \cite{addapr}.

 At high energies and small angles the analyzing power (\ref{anf})
can rewritten  in form
\ba
 - A_N\frac{d\sigma}{dt}/2& =&
   ImF_{nf}^h (ReF_{sf}^c +  ReF_{sf}^{h}) +
Im F^{c}_{nf} (Re F^{c}_{sf}  + Re F^{h}_{sf})  \\ \nonumber
   &&  - ImF_{sf}^c ( Re F_{nf}^{c} + Re F_{nf}^{h})
     - ImF_{sf}^h (ReF_{nf}^c  + ReF_{nf}^h ).    \\ \nonumber
\ea

 We obtain for proton-proton scattering at high energies at the point  $t_{re}$
 where $Re F^{nf}_h = - Re F^{nf}_c$
\ba
  Re F_{sf}^{h}(s,t) = \frac{-1}{2  (Im F_{nf}^{h}(s,t)+Im F_{nf}^{c}(t))}
  A_{N}(s,t) \
    \frac{d\sigma}{dt} \
	       - Re F_{sf}^{c}(t).   \lab{refhm}
\ea
 We can again take the hadron spin-nonflip
 and spin-flip
  amplitudes with  definite parameters and calculate the magnitude of
 $A_N$ by the usual complete form (\ref{anf})
 while  the real part of the hadron
 spin-flip amplitude is given by (\ref{refhm}).
 Our calculation by this formula and the input real part of the
  spin-flip amplitude are shown in
fig. 3.
  At the point $t_{re}$ both curve coincides. So if we obtain from
 the accurate measurement  of the differential
 cross sections the value of $t_{re}$, we can find from
   $A_N$ the value of the real part of the hadron spin-flip
 amplitude at the same point of momentum transfer.


\begin{figure}
\epsfysize=8.cm
\epsfxsize=12.cm
\centerline{\epsfbox{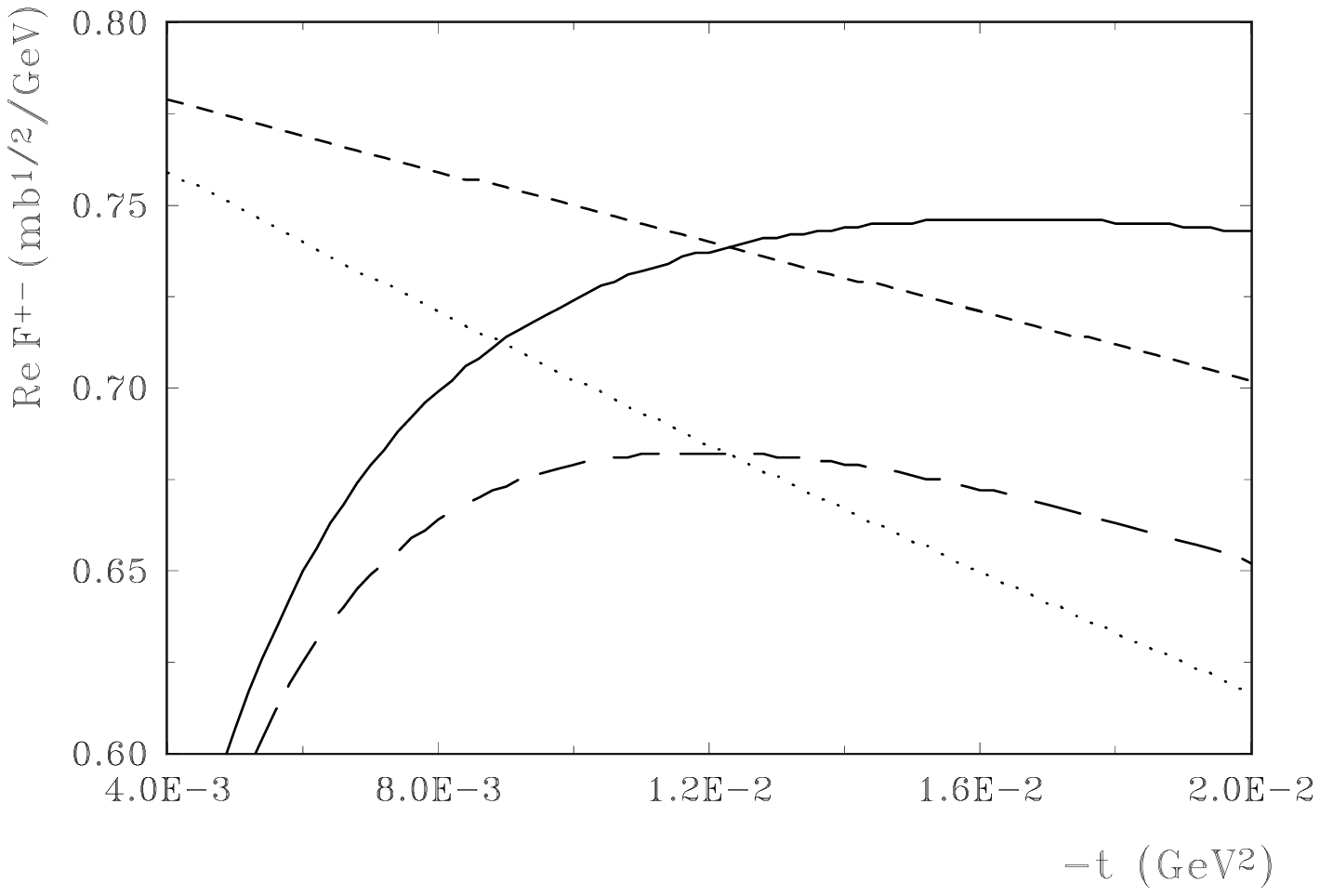}}

\vspace*{-5mm}

 Fig.3. \hspace{2mm}{ The form of $Re ( F^{sf})$ }:
 solid and long-dashed lines are calculations by (14); \\
\phantom{.} \hspace{2.cm} short-dashed  and
  dottes lines are model amplitudes \\
\phantom{.} \hspace{2.cm} with the slopes $B_{1}^{-} = B_{1}^{+}$ and
  $B_{1}^{-} = 2 \ B_{1}^{+}$.
\end{figure}


\section{Conclusion}

  By  accurate measurements of the analyzing power in
   the  Coulomb-hadron
   interference region we can find the structure of the hadron spin-flip
   amplitude, and this gives us further information about the
  behavior of the hadron interaction potential at large distances.
      Such contribution can be taken into account in the peripheral
  dynamic model \ci{zpc}.
 In fact, the model takes into account the contribution of the hadron
 interaction at large  distances, and
 the calculated hadron spin flip amplitude
  leads to the  ratio of the slopes of the ``residual'' spin-flip and
  spin-nonflip amplitudes at small momenta transfer.
 The model gives also the large spin effects in the diffraction dip
  domain \cite{yaf-str}.
  We should note that all our consideration are based
  on the usual supposition that
  the imaginary part of the high energy scattering amplitude
  has an exponential behaviour.
  The other possibility, that  the slope changes slightly
   with  $ t \rightarrow 0 $, requires a more refined discussion
   that will be the subject of a subsequent paper.

\vspace*{5mm}

 {\it Acknowledgements.} One of us (OVS)  would like to thank
  the Department of Theoretical Physics of the University of Torino for
 its hospitality and the INFN for its financial support.


\end{document}